\documentclass[pra,aps,amssymb,amsfonts,amsmath,showpacs,twocolumn]{revtex4}
\pdfoutput=1

\usepackage{graphicx}
\usepackage{ucs} 

\usepackage{color} 
\definecolor{darkblue}{rgb}{0,0,0.5}
\definecolor{lila}{rgb}{0.3,0,0.3}
\definecolor{turq}{rgb}{0,0.1,0.4}

\usepackage{url} 

\usepackage[pdftex,
colorlinks=true,
 linkcolor=darkblue, 
 filecolor=red,
 citecolor=turq, 
 urlcolor=lila, 
 pdftitle={Coherent State Preparation and Observation of Rabi Oscillations in a Single Molecule},
 pdfauthor={Ilja Gerhardt, Gert Wrigge, Gert Zumofen, Jaesuk Hwang, Alois Renn, Vahid Sandoghdar},
 pdfsubject={Coherent State Preparation of a Single Molecule},
 pdfkeywords={Single Molecules, Resonance Fluorescence, Fluorescence Spectroscopy, Extinction Spectroscopy, Extinction, Absorption, Solid Immersion Lens, SIL, Homodyning, Coherent State Preparation, Rabi Oscillations},
 pdfpagelabels=true,
 breaklinks=true,
 plainpages=false,
 bookmarks=false, bookmarksnumbered=true
]{hyperref}

\renewcommand{\d}{{\rm d}}
\newcommand{\Om}{{\mathit \Omega}}

\begin{document}

\title{Coherent State Preparation and Observation of Rabi Oscillations in a Single Molecule}

\author{I. Gerhardt}\altaffiliation{present address: CQT, Centre for Quantum Technologies, National University of Singapore, 3 Science Drive 2, 117543 Singapore}\email{ilja@quantumlah.org}
\author{G. Wrigge}
\author{G. Zumofen}
\author{J. Hwang}
\author{A. Renn}
\author{V. Sandoghdar}
\affiliation{Laboratory of Physical Chemistry, ETH Zurich, CH-8093 Zurich, Switzerland}

\begin{abstract}

We report on the excitation of single molecules via narrow zero-phonon transitions using short laser pulses. By monitoring the Stokes-shifted fluorescence, we studied the excited state population as a function of the delay time, laser intensity, and frequency detuning. A $\pi$-pulse excitation was demonstrated with merely 500 photons, and 5 Rabi cycles were achieved at higher excitation powers. Our findings are in good agreement with theoretical calculations and provide a first step toward coherent manipulation of the electronic states of single molecules with few photons.

\end{abstract}


\maketitle

Laboratory preparation of a system in a well defined quantum mechanical state is central to many activities in physics and chemistry. An important example of current interest is in quantum information processing, where methods are sought to prepare, store, and read out qubits. An attractive scheme for achieving these goals is to use photons as information carriers because they can be transmitted over large distances in a robust manner~\cite{bouwmeester:00}. Also in chemistry, manipulation of the quantum states of molecules by light has been vigorously pursued. For instance, molecular interactions on the time scales of vibrational oscillations have been studied in ensembles~\cite{Gruebele:93}. However, such experiments remain very challenging for room temperature investigations at the single molecule level~\cite{vanDijk:05}.

Efficient manipulation of the electronic states of a single emitter requires a large coupling with photons. In the past, high-finesse cavities have been employed to enhance this interaction~\cite{Boozer:07}, but recent experiments have shown that substantial coupling is also possible in a single pass arrangement where a light beam is tightly focused on the emitter~\cite{Wrigge:08,Vamivakas:07,Tey:08}. In fact, our theoretical studies have shown that it should be possible for a single atom to perfectly reflect monochromatic light and therefore imprint a $\pi$ phase shift on it~\cite{Zumofen:08}. In this article, we take the first step in exploring pulsed excitation and coherent preparation of a single molecule electronic state.

Cryogenic single molecule spectroscopy takes advantage of the inhomogeneous distribution of the molecular resonances in a solid~\cite{SMbook}. Typical dopings of about $10^{-6}$ yield one molecule in a focused laser beam within a spectral region of few GHz. By scanning the frequency of the laser, one can selectively tune to the zero-phonon lines (ZPL) of individual dye molecules, with life-time limited linewidths down to 10-50~MHz. While selection of single molecules is straightforward with narrow-band tunable lasers, it is nontrivial under pulsed excitation. If the pulses are too short, several molecules could be excited at the same time. If the pulses are longer than the excited state lifetime, they are ineffective owing to the dominant damping of the state coherence by spontaneous emission. These considerations restrict the choice of the pulsed light source for coherent spectroscopy of single fluorescent molecules.

Coherent superposition of the ground and excited states of an emitter under pulsed laser excitation can be described by solving the optical Bloch equations~\cite{AllenEberly}. The dimensionless pulse area \begin{equation} A =\int_0^T  \sqrt{\Delta^2+|\mathit{\Omega}|^2}\, dt\end{equation} yields the Bloch vector nutation angle, where $T$ is the pulse duration, $\Delta$ denotes the detuning between the laser frequency $\omega$ and the transition frequency $\omega_0$, and $\mathit{\Omega}(t)=-d_{12}\mathcal{E}(t)e^{i \varphi(t)}/\hbar$ is the complex Rabi frequency. Here $d_{12}$ represents the transition dipole moment, which we assume to be parallel to the incident electric field. The real-valued quantity $\mathcal{E}(t)$ is the electric field amplitude that characterizes the pulse shape, and $\varphi(t)$ signifies a time-dependent phase. A deterministic preparation of the emitter in its upper state can be accomplished by the proper adjustment of $A$. Rabi flopping in atomic ensembles~\cite{Gibbs:73}, single semiconductor quantum dots \citep{Stievater:01,Htoon:02}, as well as single trapped atoms \citep{Darquie:05}, and ions~\cite{Roos:99} have been successfully observed in this fashion. In the case of single dye molecules, Rabi oscillations have been explored via intensity autocorrelation measurements performed in start-stop continuous-wave (CW) illumination~\citep{Basche:92b,Wrigge:08}, and by rf-Stark shift sweeping of $\Delta$~\cite{Brunel:99}.

\begin{figure}[b!]
\centering
\includegraphics[width=7.0cm]{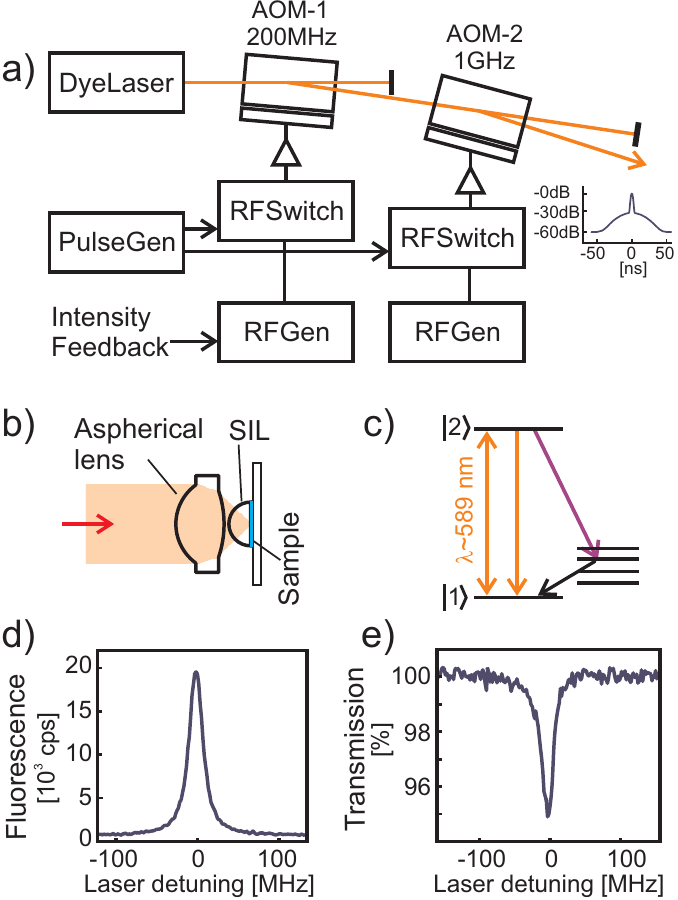}
\caption{a) The arrangement of two cascaded acousto-optical modulators to obtain short laser pulses with a high signal to background ratio. b) The optical setup inside the cryostat with a solid immersion lens (SIL). c) The level scheme of a single molecule. d), e) Typical examples of CW fluorescence and transmission spectra, respectively. \label{setup}}
\end{figure}

In this work we studied the dye molecule dibenzantanthrene (DBATT) embedded in a {\emph n}-tetradecane Shpol'ski\unichar{0301}-matrix. The excited state in DBATT has a fluorescence lifetime of $T_1=9.5$~ns~\cite{Lettow:07}, corresponding to a radiatively broadened linewidth of $\mathit{\Gamma}_1/2\pi= 17$~MHz for the ZPL transition. We used a CW single mode dye laser (Coherent 899
Autoscan, $\Delta \nu \leq 1\,{\rm MHz}$) to identify single molecules via fluorescence excitation spectroscopy~\cite{Orrit:90}. As sketched in Fig.~\ref{setup}b, a confocal microscope based on a solid immersion lens~\cite{Wrigge:08} enabled us to efficiently excite the molecules and detect them both via their Stokes-shifted fluorescence and via extinction spectroscopy (see
Figs.~\ref{setup}d, e). The red shifted fluorescence was filtered from the excitation laser light by an optical long pass filter and was detected using an avalanche photodiode (APD).

To generate laser pulses, we used a cascade of two acousto-optical modulators (AOM-1, 200~MHz and AOM-2, 1~GHz) to chop the output of the CW laser (see Fig.~\ref{setup}a). A pulse generator (SRS, DG535) controlled the rf-switches (Minicircuits) located between the oscillators and the output stages of the drivers (Brimrose). Concentric pulses were generated with pulse times of 50~ns for AOM-1
and 2.9~ns or longer for AOM-2, resulting in a sharp pulse riding on top of a broader pedestal. Both AOMs were operated in the Bragg regime~\cite{bornwolf}, and the peak to background intensity attenuation was approximately -30~dB for each device. The pulse repetition rate was set to 700~kHz, corresponding to a period of about 1.4~$\mu s$. We verified that the average intensity decreased
linearly and extrapolated to a negligible value as we lowered the repetition rate to zero. The pulse intensity was measured by monitoring the optical output on a fast silicon PIN diode. Additionally, we used an integrating photodiode to give a reference for the average incident power, which was stabilized by feedback to the modulation input of the AOM-1 driver.

\begin{figure}[t!]
\centering
\includegraphics[width=7.0cm]{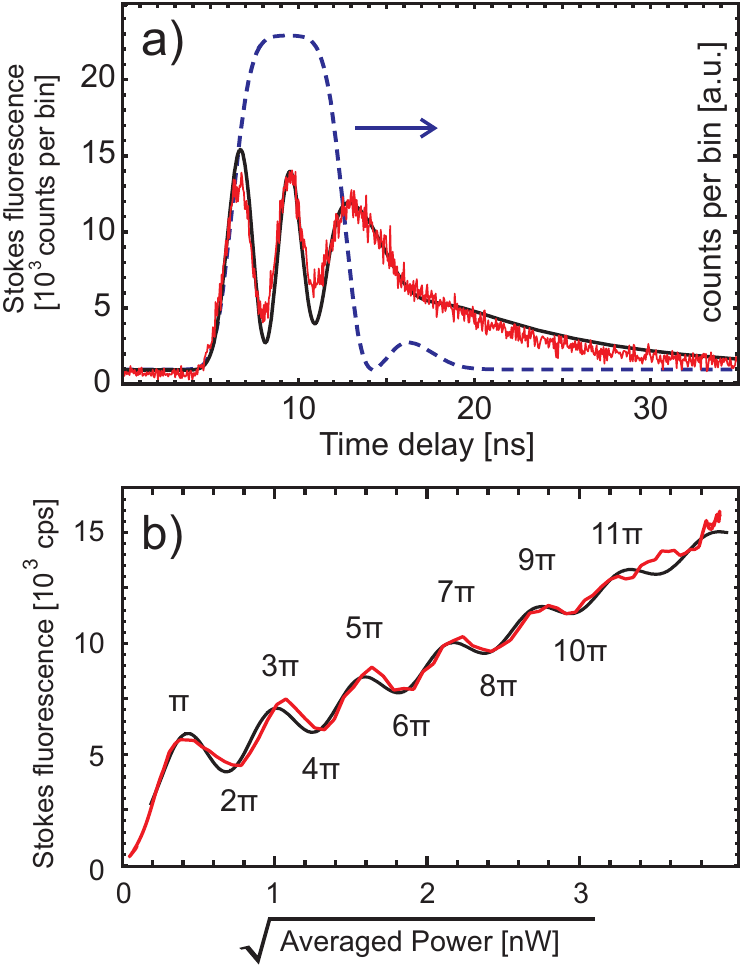}
\caption{Raw Stokes-shifted fluorescence of a single molecule (red curve) and a theoretical fit (black curve): a) as a function of delay with respect to the optical pulse. The blue dashed line displays the measured excitation pulse intensity. b) as a function of the square root of the excitation pulse power using an optical pulse of 4 ns length. The data correspond to the vertical cut at $\Delta=70$ MHz in Fig.~\ref{many-Rabi}a. \label{two-Rabi}}
\end{figure}

The red curve in Fig.~\ref{two-Rabi}a, shows two full Rabi cycles recorded during the passage of a long pulse on resonance and the subsequent decay of the excited state population caused by spontaneous emission. Here, the output of the pulse generator that controlled the AOMs was attached to a time-amplitude converter (Becker \& Hickl, SPC-130) and triggered the start of a time correlation measurement. Upon detection of a fluorescence photon on a fast APD (MPD, $\approx$50~ps jitter), the measurement was stopped and the corresponding arrival time was recorded. We also analyzed the excitation pulses using the APD in the same fashion as the fluorescence photons by using a mirror to reflect some of it. The blue dashed line in Fig.~\ref{two-Rabi}a shows the excitation pulse
with a high time resolution. These measurements reveal a small after pulse that perturbs the exponential decay of the excited state (see Fig.~\ref{two-Rabi}a).

Because the pulse width is of the same order of magnitude as $T_1$, the molecule has a certain probability of getting re-excited within the pulse duration. Nevertheless, our calculations show that the detected fluorescence signal is, to a very good approximation, proportional to the population of the excited state at each delay time. This follows because first, one has to account for a 70~ns
dead time of the APD, which limits the detection to at best one photon per pulse. Moreover, because the overall photon detection efficiency is only a few per cent, the histogram of photon arrival times does not represent the probability density of the first emitted but of the first detected photon~\cite{he_physchemchemphys_2006}. To compare our results with theory, we  used the independently determined shape of ${\cal E}(t)$ and solved the time dependent Bloch equations for a two-level atom numerically. The solid black curve in Fig.~\ref{two-Rabi}a shows the outcome fitted to the experimental data, yielding a very good agreement. We thus deduce a maximum Rabi frequency of $\mathit{\Omega}_{\rm max}/2\pi \simeq$~370~MHz and a pulse area of $A \simeq 5.7~\pi$.

Another option for monitoring Rabi flopping is to vary $\mathit \Omega$. Fig.~\ref{two-Rabi}b displays the result of an experiment where we have fixed the pulse duration to 4~ns and increased the laser intensity. The averaged fluorescence on an APD (Perkin-Elmer) with 100~ms integration time reveals more than 10$\pi$ oscillations or equivalently 5 cycles of excitation and stimulated emission. To
our knowledge, this constitutes the largest number of optical Rabi oscillations observed to date from a single emitter in the solid state. Moreover, considering a pulse period of 1.4~$\mu s$, a pulse width of 4~ns, and the average power shown in Fig.~\ref{two-Rabi}b for achieving a $\pi$-pulse excitation, we conclude that we have inverted the population of a molecule with only 500 laser photons.
This is a promising step toward excitation of an emitter with a few or even single photons~\cite{Zumofen:08,Stobin'ska:08}. 

The oscillations in Fig.~\ref{two-Rabi}b do not show the full modulation depth because the pulse duration $T$ is comparable to $T_1$. Furthermore, we observe a gradual decay of these oscillations as the excitation power is increased. We attribute this wash out effect to small fluctuations $\sigma_A$ in the pulse area $A = \overline{\mathit{ \Omega}} T$, where we have introduced the average
Rabi frequency $\overline \Om$. The AOM device specifications indicate pulse rise time fluctuations of about 200~ps. If we account for the same level of drop-off fluctuations, we find a relative variation $\sigma_T /T \simeq 7\,\%$. It thus follows that $\sigma_A$ increases linearly with the field strength. As confirmed by our simulation calculations, this results in a gradual decay of
the oscillation amplitudes in very good agreement with our experimental observations. We note that the influence of fluctuations in $\overline \Om$ were negligible in our experiment. Finally, the Rabi oscillations in Fig.~\ref{two-Rabi}b ride on a background. The black curve in this figure presents a sinusoidal fit to the experimental data, assuming linear dependencies on the
excitation field, the modulation depth reduction, and on the background increase. We remark that backgrounds have also been encountered in measurements of Rabi oscillations on quantum dot, but have been usually subtracted in the presentation of the data.

\begin{figure}[b!]
\centering
\includegraphics[width=8 cm]{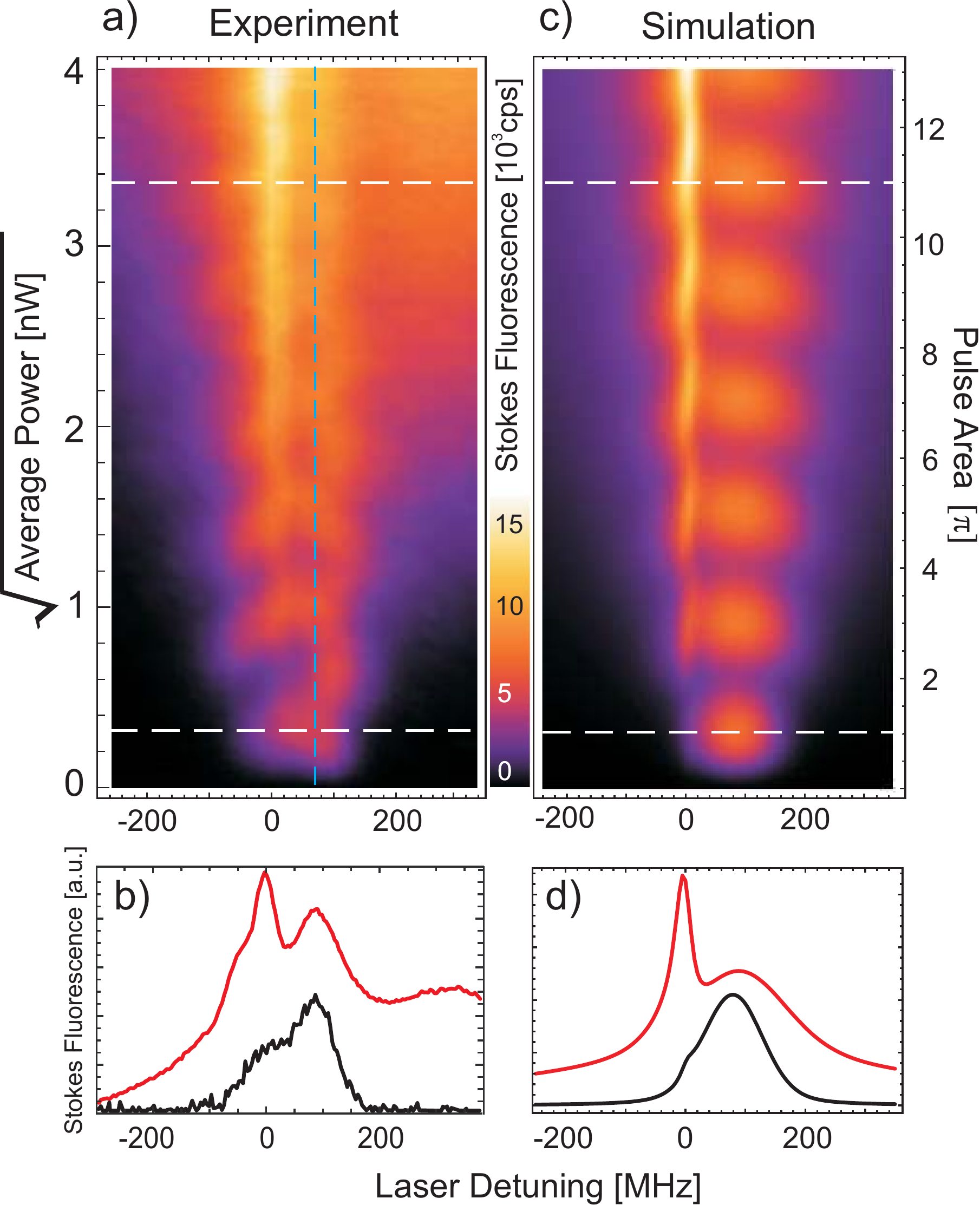}
\caption{a) Time-averaged raw fluorescence signal of a single molecule under a 4~ns pulsed excitation as a function of the excitation laser frequency (horizontal axis) and electric field (vertical axis). The vertical dashed line indicates the cross section shown in Fig. ~\ref{two-Rabi}b). c) Theoretical simulation of the data in a). b) and d) are spectral profiles taken at field strengths marked by the horizontal dashed lines in panels a) and c), respectively. \label{many-Rabi}}
\end{figure}

A third parameter that can influence the pulse area is the laser frequency detuning $\Delta$. Figure~\ref{many-Rabi}a displays a two-dimensional density plot of the fluorescence signal as a function of $\Delta$ and $\mathit \Omega$. Two cross sections at low (black) and high (red) excitation intensities are plotted in Fig.~\ref{many-Rabi}b. While both parameters contribute to $A$ in a
similar manner (see Eq.~(1)), it has to be kept in mind that a large $\Delta$ results in a less effective excitation and a diminishing fluorescence signal. Therefore, one would expect a series of concentric arches centered at $\Delta={\mathit \Omega}= 0$ in Fig.~\ref{many-Rabi}a. However, as seen in Figs.~\ref{many-Rabi}a,b, the situation is complicated.

In order to gain insight into the observed spectra, we now turn to the outcome of simulations. We have considered an excitation field ${\cal E}(t)={\cal E}_1(t)+{\cal E}_2(t) e^{i \varphi_2}$ as the sum of the contributions of AOM-1 and AOM-2. We took $|{\cal E}_{1,2}(t)|^2$ to be concentric Gaussians with widths of 50 ns for AOM-1 and 4~ns for AOM-2 and fixed the intensity ratio to 34~dB
according to our independent characterization of these instrumental parameters. Furthermore, we assumed a linear time dependence for the phase $\varphi_2(t)$ to obtain a frequency shift of $|\d \varphi_2/\d t| = 70$~MHz as observed by peaks in Figs.~\ref{many-Rabi}a,b at this detuning. Figures~\ref{many-Rabi}c and d show the simulated counterparts of Figs.~\ref{many-Rabi}a and
b. The good qualitative agreement between the experimental and theoretical data suggests the following conclusions: i) The narrow peak at zero detuning signifies fluorescence excitation by ${\cal E}_1$ in the time window when AOM-2 is off. ii) In these periods ${\cal E}_1$ imposes a background that increases approximately linearly with the field strength. In fact, the pulse areas $A_{1,2}$
obtained from the integration of the fields ${\cal E}_{1,2}(t)$ turn out to be similar. iii) At low laser intensity, the spectrum is dominated by ${\cal E}_2 e^{i \varphi_2} $. Moreover, our simulation calculations based on a volume-scattering technique \citep{maydan_ieeejqe_1970} indicate that a considerable frequency chirp can be caused by the interaction of a short acoustic wave
packet with the Gaussian laser beam in AOM-2. The short duration of this field and its chirp lead to a structure that is broad and blue shifted by approximately 70~MHz. iv) The broad shoulder at $\Delta \simeq$ 300 MHz (see Fig.~\ref{many-Rabi}b) is most probably due to the zeroth-order deflection of AOM-1 in combination with the chirp of AOM-2.

In conclusion, we have shown that well defined superpositions of the ground and excited states of a single molecule can be prepared by using pulsed laser excitation. Furthermore, we have mapped the system response as a function of laser frequency detuning and intensity. In these first experiments the pulses were generated by acousto-optical modulators and were accompanied by spurious effects,
which we analyzed in some detail. Use of higher quality pulses will substantially improve the results, paving the way for efficient coherent control and pump-probe experiments at the single molecule level. In particular, the dynamics of phenomena such as triplet state transitions, spectral diffusion, and dephasing in a solid matrix can be studied in a time-resolved fashion. The techniques
presented in this work can also be used for the preparation and manipulation of entangled states, for instance in coupled emitters~\cite{Hettich:02}. In the current work, we showed that strong focusing allowed us to achieve a $\pi$-pulse with only 500 pump photons and demonstrated 5 full Rabi cycles. In future, we will explore the prospects of single molecule excitation with very few or even single photons.

IG and VS acknowledge fruitful discussions with Ch. Kurtsiefer and B. Lounis respectively, regarding the generation of laser pulses. This work was financed by the Schweizerische Nationalfond (SNF) and the ETH Zurich initiative for Quantum Systems for Information Technology (QSIT).


\end{document}